\documentclass[aip,jap,twocolumn,groupedaddress]{revtex4}

\usepackage{amsfonts} 
\usepackage{amssymb}  
\usepackage{graphicx} 
\usepackage{graphics}
\usepackage[dvips,final]{epsfig}  
\usepackage{amsmath}  
\usepackage[latin1]{inputenc}
\usepackage{latexsym}
\usepackage[amssymb]{SIunits}

\begin{document}

\title{Upstream contamination in water pouring}

\author{S. Bianchini, A. Lage-Castellanos and E. Altshuler}
\affiliation{"Henri Poincar\'{e}" Group of Complex Systems, Physics Faculty,
University of Havana, 10400 Havana, Cuba}

\date{\today}

\begin{abstract}
We report the observation of upstream transport of floating particles
when clear water is poured on the surface of a flat
water surface on which mate or chalk particles are sprinkled. As a result,
particles originally located only at the surface of the
lower container can contaminate the
upper water source by ``riding" on vorticial water currents. We speculate
that Marangoni forces in combination with geometry-induced vortices
may explain the observed phenomenon.\\[0 cm]
\end{abstract}

\pacs{???}

\maketitle


We briefly report a phenomenon apparently not previously reported in the
literature, which may have potential implications
in many chemical, medical, pharmaceutical and industrial processes.

The phenomenon was first observed during the preparation of the typical
Argentinian drink, {\it mate}, when hot water was poured, from a pot,
on a water surface ``contaminated" with floating mate particles (each
particle is like a grass leave of an average area near $ 0.5 mm^2$).
If the column of falling water was short enough (say, under $1 cm$-height),
particles of mate were observed to ``swim up the stream", actually reaching the
originally ``uncontaminated" water pot.

We checked that the
phenomenon could take place with the two water containers at room temperature,
and that chalk particles (instead of mate leaves) worked as well.

\begin{figure}[b!!]
  \includegraphics[height=7.0cm, width=8.0cm]{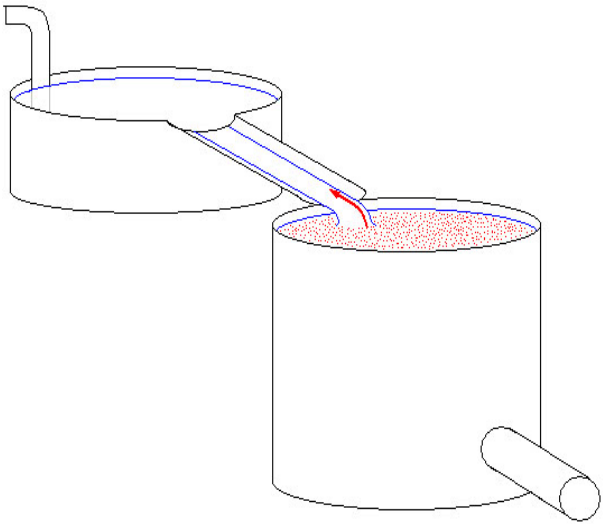}
  \caption{Experimental setup. As clean water is poured from the upper to the lower container through a horizontal channel, particles added to the surface of the lower container eventually ``climb up the stream" and contaminate the channel (as illustrated by the arrow). The water level at both containers is kept constant. }\label{setupvertical}\vspace{-0.2cm}
\end{figure}

\begin{figure}[b!!]
  \includegraphics[height=4.0cm, width=8.0cm]{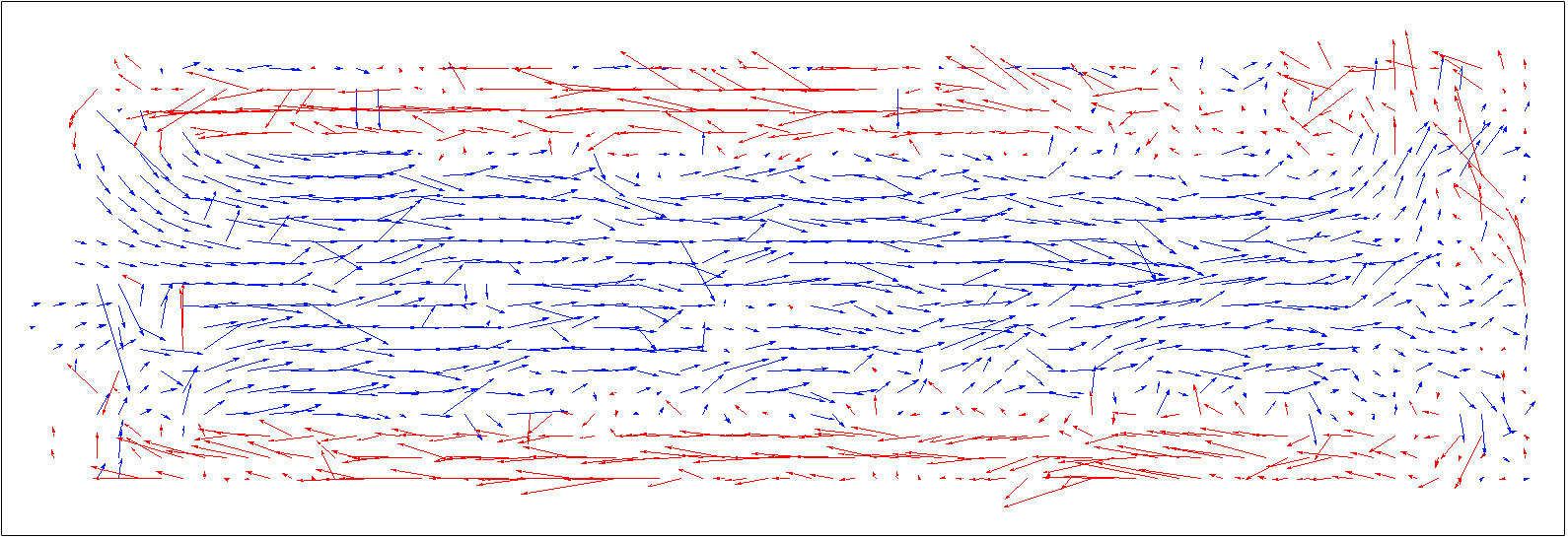}
  \caption{Velocity field of the floating particles on the surface of the channel, after ``climbing up" the stream from the lower container. The upper container feeds the channel from the left, which discharges to the lower container at the right. The red arrows correspond to velocities with a component against the main flow of water: they are located near the lateral walls of the channel.}\label{verticalchannelvortices}\vspace{-0.2cm}
\end{figure}

In order to perform controlled experiments, we constructed a ``pouring setup" illustrated in Figure~\ref{setupvertical}.
A flux of water
in the range $0$-$15$ cm$^3/$s flows along a $62$-mm-long horizontal channel
shaped as a half cylinder of $20$ mm radius. As the water reaches the edge of the
channel, it falls on a large container of water,
whose surface is at a distance $h$ underneath the
edge of the channel (all parts are made of aluminium).
For distances of the order of $1$ cm or less,
some of the floating particles eventually start to
``climb up the stream" and invade the channel, as part of one of the two
elongated vortices appearing along the two sides of the channel .
The vorticity is
such that the upstream particles flow near the edges of the channel, while the
downstream particles flow near the center of the channel (see Fig~\ref{verticalchannelvortices}).
Apparently, the vortices continue
along the free falling stream, and reach the water surface in the lower reservoir.

\begin{figure}[b!!]
  \includegraphics[height=6.0cm, width=8.0cm]{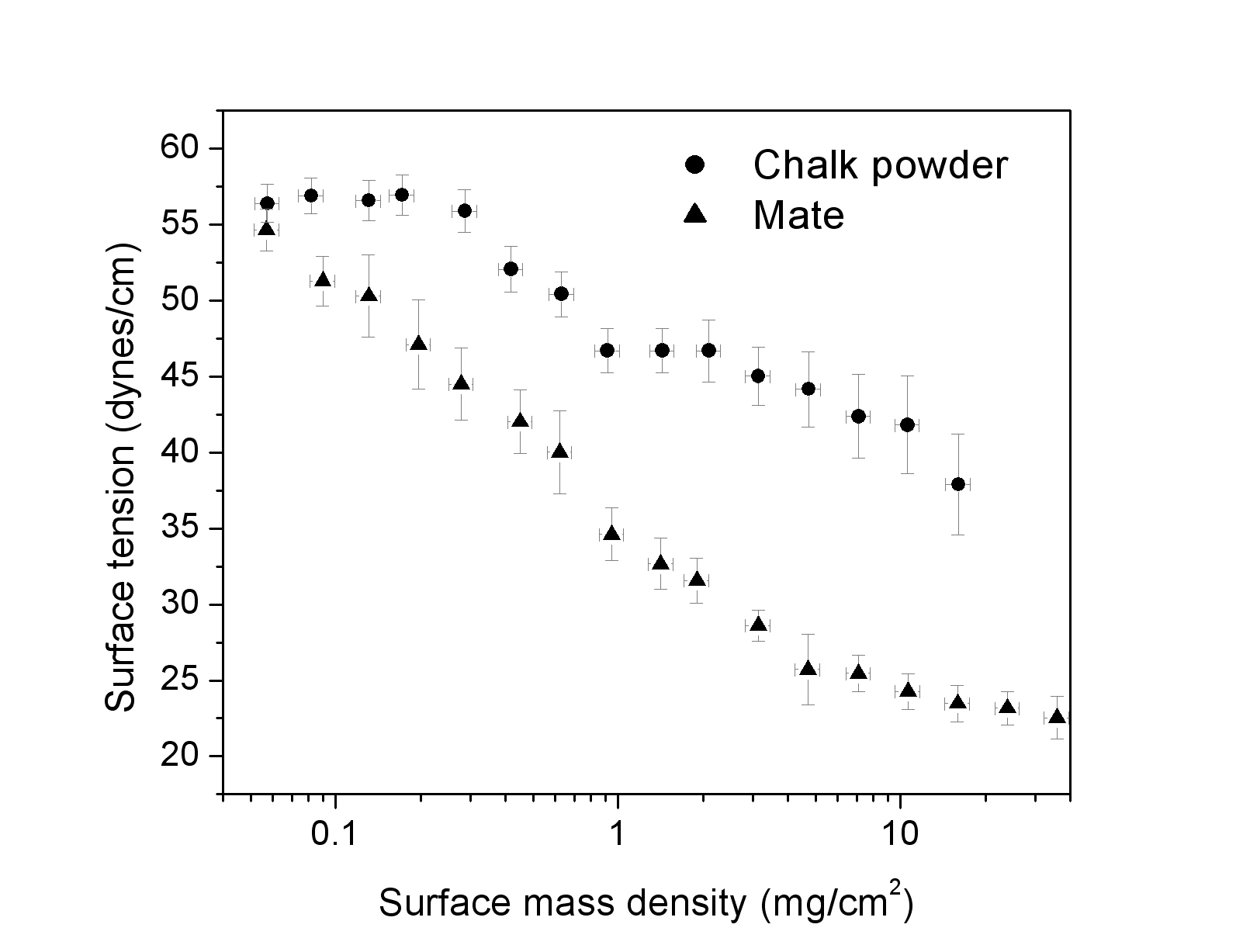}
  \caption{Dependence of the surface tension of water on the superficial density of mate and of chalk particles, measured using the ring method.}\label{SurfaceTensionMeasurements}\vspace{-0.2cm}
\end{figure}

While our vortices resemble those well-known to occur in horizontal channels due to
edge effects~\cite{LibroSotolongo},
their existence has not been reported, as far as we know, in connection to falling streams of water.
We speculate that the
``upstream" vortex formation could be reinforced by a Marangoni-type
force~\cite{SomeOtherTextBook} associated to a decrease of the
surface tension in the lower reservoir due to particle addition.
To explore this possibility, we measured the influence of
different surface densities of mate and chalk particles on the surface tension of
water, using the Ring Method~\cite{RingMethod}. As seen in Fig~\ref{SurfaceTensionMeasurements}, we have
demonstrated that the surface tension decreases as the surface density increases for both
materials.

We are currently performing experiments on a
horizontal channel connecting two water reservoirs, where a water stream is
imposed from right to left. Then, we add mate or chalk particles
to the left container, and study the vortices appearing in the channel by tracking the
floating particles. The idea is to experimentally evaluate under
what conditions such vortices can persist when as the horizontal setup ``evolves"
into the water-pouring channel geometry.

\section*{Acknowledgments}

The authors acknowledge useful discussions with J. E. Wesfreid, E. Cl\`{e}ment, D. Qu\'{e}r\'{e}, J. Gollub, O. Sotolongo-Costa and L. del R\'{i}o.
E. Altshuler and A. Lage thank R. Mulet for driving their attention to the serendipitous
findings of S. Bianchini.

\bibliographystyle{aiprev}

\end{document}